\documentclass[journal]{IEEEtran}

\usepackage{graphicx}
\graphicspath{ {./images/} }

\usepackage{lipsum}
\usepackage[dvipsnames]{xcolor}
\usepackage{textcomp}

\usepackage[utf8]{inputenc}
\usepackage[english]{babel}

\usepackage[backend=biber, style=ieee]{biblatex}
\usepackage{textgreek}

\begin{document}

\title{Frequency Response of Metal-Oxide Memristors}

\author{{Vasileios Manouras, Spyros Stathopoulos, Suresh Kumar Garlapati, Alex Serb,~\IEEEmembership{Senior Member,~IEEE} and Themis Prodromakis,~\IEEEmembership{Senior Member,~IEEE,}}
\thanks{This work was supported by the Engineering and Physical Sciences Research Council (EPSRC) under Grant EP/R024642/1 and the European Union Horizon 2020 Project, Grant No.824162.}
\thanks{All authors are affiliated with the Centre for Electronics Frontiers, Zepler Institute for Photonics and Nanoelectronics, University of Southampton, Southampton, SO17 1BJ, U.K. (e-mail: V.Manouras@soton.ac.uk; s.stathopoulos@soton.ac.uk; garlapatisureshkumar@gmail.com;  a.serb@soton.ac.uk; t.prodromakis@soton.ac.uk).}}%

\maketitle

\begin{abstract}

Memristors have been at the forefront of nanoelectronics research for the last decade, offering a valuable component to reconfigurable computing. Their attributes have been studied extensively along with applications that leverage their state-dependent programmability in a static fashion. However, practical applications of memristor-based AC circuits have been rather sparse, with only a few examples found in the literature where their use is emulated at higher frequencies. In this work, we study the behavior of metal-oxide memristors under an AC perturbation in a range of frequencies, from $10^3$ to $10^7$ Hz. Metal-oxide memristors are found to behave as RC low-pass filters and they present a variable cut-off frequency when their state is switched, thus providing a window of reconfigurability when used as filters. We further study this behaviour across distinct material systems and we show that the usable reconfigurability window of the devices can be tailored to encompass specific frequency ranges by amending the devices' capacitance. This study extends current knowledge on metal-oxide memristors by characterising their frequency dependent characteristics, providing useful insights for their use in reconfigurable AC circuits.

\end{abstract}

\begin{IEEEkeywords}
Memristor, AC, Frequency, Analogue, Reconfigurable, Impedance, RRAM
\end{IEEEkeywords}

\IEEEpeerreviewmaketitle

\section{Introduction}

\IEEEPARstart{M}{ore} than a decade has passed from the initial publication from HP labs \cite{Strukov} that linked Resistive Random Access Memory (RRAM) characteristics with Chua's definition of memristors \cite{Chua}. The solid-state implementations of memristive technologies led to a sudden increase in interest in the area of reconfigurable electronics. Publications covering this new area of interest made their appearance initially describing the resistive switching phenomenon \cite{SCHROEDER},\cite{SAWA}. Over the following years multiple aspects of these devices were thoroughly examined, such as, the impact of device area on performance \cite{Govoreanu} and the impact of using different metal oxide dielectrics \cite{Wei},\cite{Biju} as layers where resistive switching takes place. Currently prevalent theories for the mechanism underpinning resistive switching include oxygen vacancy movement \cite{Yang}\cite{Lai} which leads to valence changes in the active layer material \cite{Valov},\cite{Pan},\cite{Waser}. Up to this point the majority of research carried out centered around the applications of resistive switching systems in static conditions, in circuits with direct current (DC) stimuli.

\begin{figure}[h]
    \centering
    \includegraphics[width = 9cm, height = 10cm]{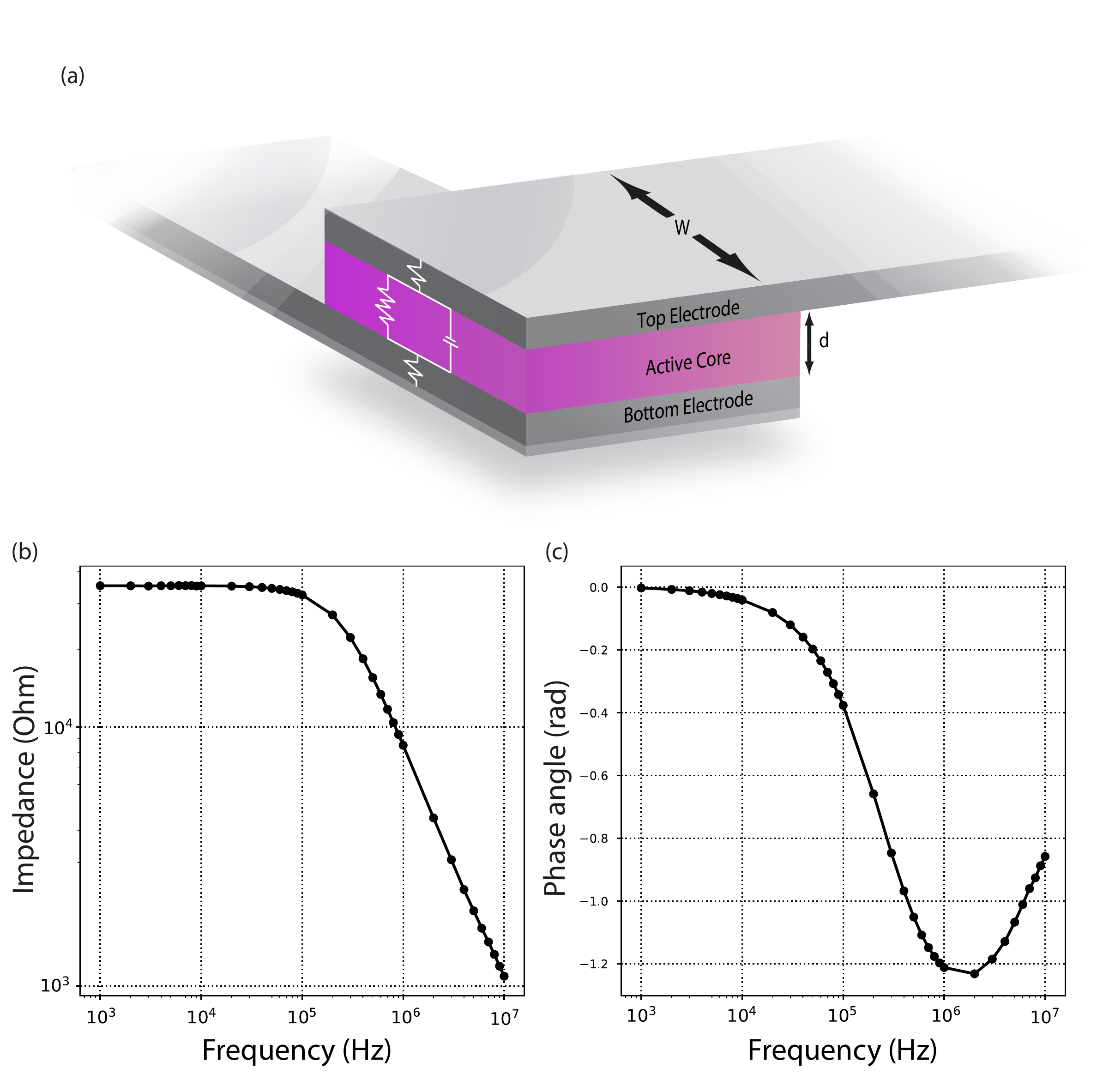}
    \caption{a) Metal oxide memristor structure. Equivalent electrical model of the device is depicted at the left side. Physical parameters which affect the equivalent model are depicted on the right side. Namely, W for electrode width and d for active core thickness. Active core can be any oxide material or layers of materials, which produce a memristive device.
    b) Initial impedance bode plot of devices, c) Initial phase bode plot of tested devices}
    \label{fig: 1}
\end{figure}

Circuits operating in alternating current (AC) mode could also benefit from using resistive switching devices, but up to this point publications concerning the effect of Metal Oxide memristor frequency depedencies were mainly centered around transient analysis of devices \cite{Mazady}, measurements in singular frequencies \cite{Yan} or as a way to characterize the conduction mechanisms of devices \cite{Lee}. At the same time, however, several publications emerged on simulating AC circuits with memristive devices \cite{Wang} \cite{Rajagopal}, indicating that there is a need for characterization of metal-oxide memristor devices in the frequency spectrum. Whilst the aforementioned publications focused on the frequency response of the proposed circuits overall, they did not account for the frequency response of the individual memristor cells used, often emulating these as distinct static resistors with uniform response across the employed frequency range. This important oversight was, on one side, ascribed to the lack of existing memristor models to account for small AC signals superimposed on DC stimuli as well as the researchers focus that in the interest to show how such tunable resistive components can tweak the response of an AC circuit, only accounted for distinct static resistive loads. 

In this work, we utilise standard electron device testing methodologies and adopt these accordingly for use in memristors. We then employ this for studying the frequency response of metal oxide memristors that are found to show a predominantly resistive behavior up to a certain frequency, but they are not equally as stable in the entirety of the frequency spectrum investigated herein. We further show that such devices are much more than static tunable resistors and need to be represented via an RC empirical model, as depicted in Figure 1(a). Variations of this electrical model have been hinted in the literature for memristors with different characteristics \cite{Lee}, \cite{Park},  \cite{DASH}, but, to the best of our knowledge, a full RC-level description with regards to the devices' switching remained outstanding till this work. We further present an analysis of the devices' behavior in a range of frequencies and assess the electrical reconfigurability of the devices over the same frequency range. Finally, we expand the remit of this study by investigating a variety of memristive devices comprising different electrode areas, dielectric thicknesses and core materials and the results are discussed.

\section{Experimental Methodology}

\subsection{Device Prototyping}

All devices were fabricated on 6-inch silicon wafers, on top of which 200 nm of silicon oxide (SiO$_2$) was thermally grown. All top and bottom electrodes consist of Platinum (12 nm), deposited by electron beam evaporation. To ensure good bottom electrode adhesion a Titanium film (5 nm) was used prior the deposition of Platinum. Devices with three different types of resistive switching layers were fabricated (TiO$_x$ , TiO$_x$/Al$_2$O$_3$ , SnO$_x$), all deposited by magnetron sputtering. The deposition power for TiO$_x$, Al$_2$O$_3$ and  SnO$_x$ was respectively 2 kW, 100 W and 70 W. Gas flow rates inside the chamber were 8 sccm for O$_2$ and 35 sccm for Ar gas for the TiO$_x$ and TiO$_x$/Al$_2$O$_3$ dielectrics, whilst the ratio used for SnO$_x$ was 10 sccm Ar and 20 sccm O$_2$. Patterning of all layers was carried out by negative tone photolithography. After each deposition a lift-off process followed, with N-Methyl-2-pyrrolidone (NMP), and a gentle surface clean with O2 plasma, by Reactive Ion Etching techniques.

\subsection{Electrical characterization}

Electrical characterization of devices in DC was carried out with the in-house testing system ArC One\textsuperscript{TM} \cite{Arc}. Initially, devices were electroformed using the corresponding built-in module of this system, whereby continuous pulses of 100 μs width and an amplitude progressively ranging from 3 to 7 V were applied to the device until a partial dielectric breakdown was achieved. After electroforming, I-V curves of the devices were taken to ensure uniform working conditions. Subsequently a brief retention test was done to evaluate the short-term stability of devices, in order to avoid any negative impact from volatile behaviour, during the measurement window. Throughout this work, the read pulse amplitude was set at 0.5 V for all measurements.

After completing the standard DC characterization process, the device behavior under AC conditions was probed by using a Keithley 4200A-SCS. To probe the frequency characteristics of these devices an initial sweep would take place to ascertain the starting condition and directly connect it to the resistive state given by the ArC system. Afterwards, a switching event would be induced by applying a voltage stress to the device, by means of a voltage sweep, and the device would be submitted to a new frequency sweep. This process is repeated for each device by applying voltage sweeps ranging from 1.5 to 3 V in magnitude, with alternating polarity. Through this process, the device is switched amongst mutliple states and a new frequency scan is taken at each state. All frequency sweeps were carried out by using a 0.5 V DC bias along with a 100 mV superimposed AC stimulus. This study was carried out for frequencies ranging between $10^3$ and $10^7$ Hz.

\begin{figure}[b]
    \centering
    \includegraphics[width = 9cm, height = 7cm]{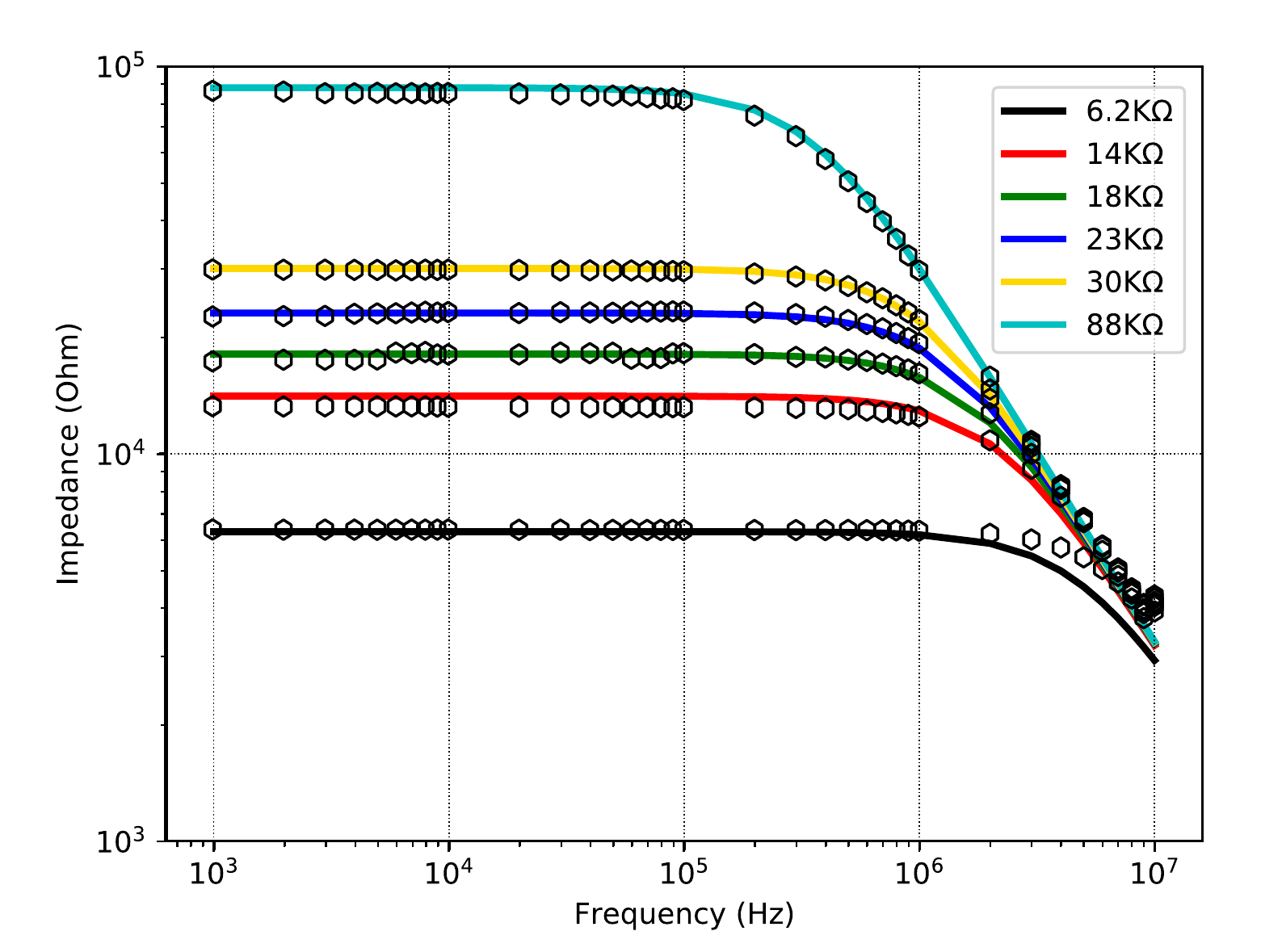}
    \caption{Device electrically programmed in different resistive states, as shown in legend. Black hexagons depict the experimental values received from testing, while the coloured lines are the simulated results for the depicted Resistances.}
    \label{fig 2}
\end{figure}

\section{Electrically programmable behavior}

There have been a few reports in the literature \cite{Lee} \cite{Park} \cite{DASH} where memristors have been described via an RC circuit. In figures 1(b,c) the initial state of a device is given with a behavior akin to that of a low pass filter. This equates to the devices showing a  stable impedance in regards to frequency, until a sharp drop occurs, defining a cut-off point. All devices tested exhibited this type of behavior thus, this type of frequency response could be extended to other devices fabricated as a Metal-Insulator-Metal (MIM) capacitor and then electroformed.
 
 \begin{figure*}[h]
    \centering
    \includegraphics[width = 16cm, height = 14cm]{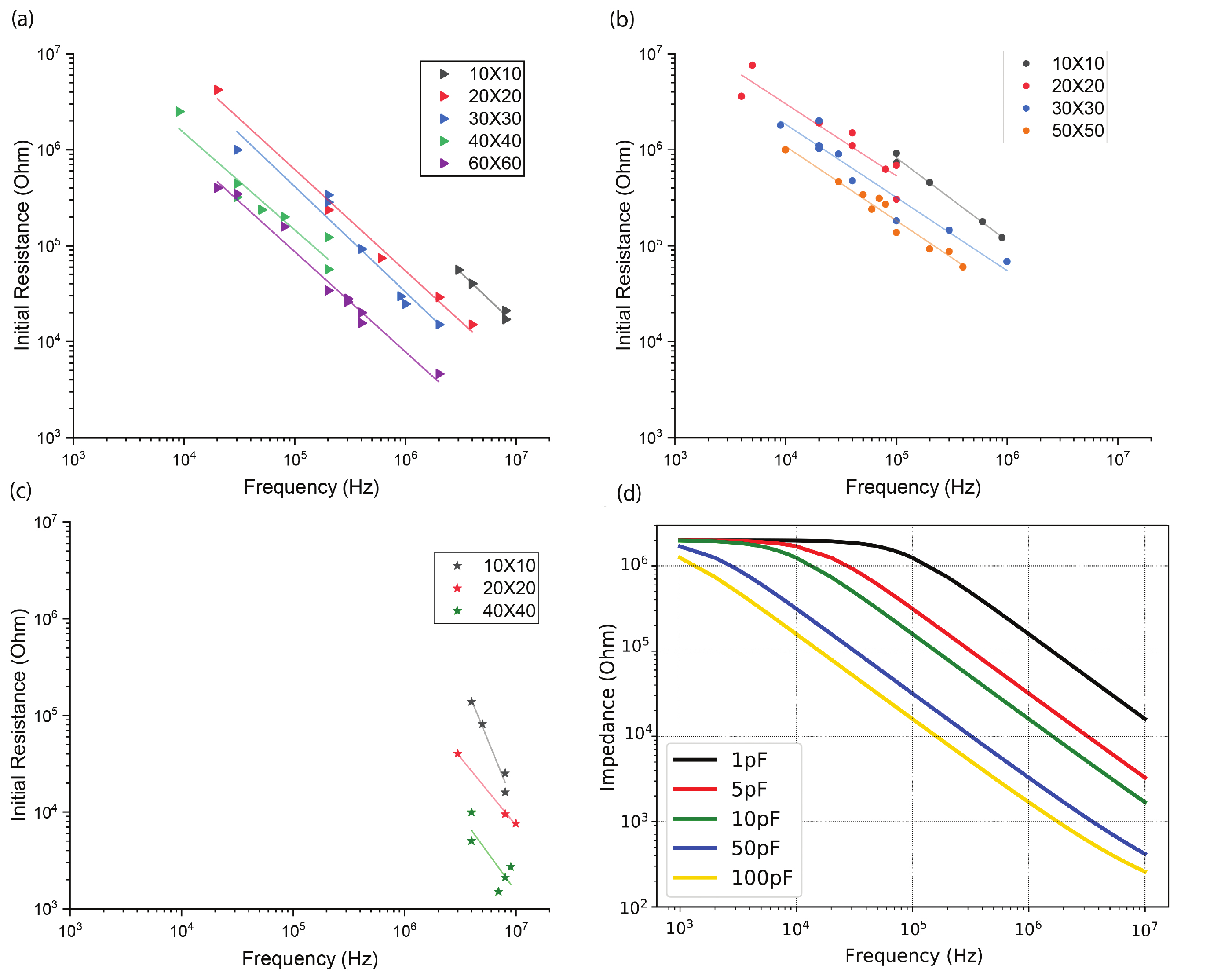}
    \caption{Tuning of programability window by changing area of device. a) Variation of electrode area in TiO$_x$/Al$_2$O$_3$ devices, b) Variation of electrode area in TiO$_x$ monolayer devices, c) Variation of electrode area in SnO$_x$ devices, d) Simulation of model depicted in Fig.1(a) with varying Capacitive values}
    \label{fig 3}
\end{figure*}

During electroforming the partial breakdown of the dielectric can lead to distinct resistive levels, depending on the forming parameters\cite{Michalas}. This process is partially controlled and guided with a conservative voltage pulsing. Throughout this process, it was found that the cut-off frequency was largely dependent on the registered resistive state of the device. Due to the nature of electroforming, damage to the dielectric is inevitable and small differences may arise between the capacitive values of different devices. Analysis of the data extracted from measurements of the devices' capacitive and resistive parameters showed a rather small device-to-device variability in capacitive values, demonstrating an overall good uniformity. On the other hand, resistive variability across all measured devices is found to be higher and thus constitutes the main driving force behind changes observed in cut-off frequency. It is important to note that it was found that line resistance had a minimum impact on the devices' switching behavior.
 
By adhering to the resistive switching protocol discussed in section II.B it was possible to measure the device response in different resistive states. The incremental and symmetrical increase of switching voltages also enabled us to factor-in the multi-bit functionality of some types of devices \cite{Stathopoulos}. These devices tested herein, switch in a non-volatile manner that was confirmed via a short retention test performed by the ArC One capability. Figure 2 depicts the impedance frequency response of a device for distinct memory-resistive states. The majority of devices tested herein, were switchable when stimulated with potentials outside the zone of -1.5 to 1.5 V.

The cut-off frequency, identified in Figure 2, follows the slope with each subsequent switching event. This suggests that the dielectric undergoes a slight modification, which is not however sufficient for creating an observable change in capacitance. On the other hand, the resistive switching results in altering the cut-off, as expected when changing the resistive value of an RC. This is depicted in simulated results overlaid as lines on top of the switching points in Figure 2, whereby plotting the impedance response of this circuit and changing the material resistance, by using the one received from measuring the device, leads to a change in cut-off frequency. Having established that metal oxide resistive switching devices posses an intrinsic low-pass filter response, it is now also clear that they can be programmed to have a variable cut-off frequency by changing the resistance of the device, thus single devices acting as tunable low pass filters. 

Whilst the dynamic range of switching for any device is defined by its OFF/ON ratio, this ratio can be tuned by appropriately selecting the physical dimensions of devices (W and d) as well as the active core material. The DC characteristics of memristive devices do not seem to be heavily impacted by small changes in device size and thickness. Nonetheless, altering these parameters can have a substantial impact on the cut-off programmability window across the frequency axis. This enables optimising the development of variable low-pass filters for a range of frequencies, and broadening the field of possible applications where such a component can be used for trimming. To test the validity of this assumption and to partially control capacitance in devices, we have expanded the scope of this study to include changes in such key parameters. Section IV centers around testing of fabricated devices, categorized by parameter changed, namely, electrode area, dielectric material and dielectric thickness.

These resistive switching devices are initially fabricated as MIM capacitors and they obtain their memristive behaviour through electroforming; the characteristics defining their correspondent capacitance are thus still relevant post-electroforming. This assumption stems from the underpinning switching mechanism of such devices. A change in valence brought about by oxygen vacancy movement should not short or otherwise alter the capacitive characteristics of the device. The only change the devices sustain is the initial electroforming, which, although known to be destructive, does not result in a complete breakdown of the active layer, as shown in Figure 1(b,c). To check the validity of this assumption, key parameters were changed, such as, active layer (TiO$_x$ , TiO$_x$/Al$_2$O$_3$  and SnO$_x$), electrode area (a range from 10$\times$10 μm$^2$ up to 60$\times$60 μm$^2$) and dielectric thickness (15 nm and 25 nm).

\section{Impact of device characteristics}

\subsection{Variation of device area}

Devices were fabricated with an assortment of different electrode areas, from 10$\times$10um to 60$\times$60um, effectively defining distinct effective capacitance values per device. Initial tests showed that devices with distinct capacitance values, exhibit similar initial resistive states and electric programmability characteristics, as mentioned in section III. By increasing or decreasing capacitance the entire spectrum is transported parallel to the frequency axis. This is evident in Figure 3(a), where the cumulative initial resistances and cut-off frequencies for each electrode area subset of the TiO$_x$/Al$_2$O$_3$ devices, form the slope whose range ultimately represents the programmability window of said devices. Its has already been established that the variable cut-off frequency of each device must fall on this line. This behavior is consistent amongst devices of different size, and we further show that by scaling the area it is also possible to move this slope. This is also in excellent agreement with our simulated results, Figure 3(d), where the same equivalent model is employed, whilst only amending the corresponding capacitance value dictacted by the material of choice.

In the case of bilayer prototype memristors, the resistive state of each device under test was not found to impact their cut-off slope. On the other hand, single layer , TiO$_x$, devices exhibit this behavior only for devices formed inside a specific resistance window, as shown in Figure 3(b). When formed in lower resistive states, their slope becomes steeper. This behavior is further analyzed in section IV.B . Lastly, all devices with a SnO$_x$ switching layer appear to act in a comparable fashion, yet, these tend to form only in low resistive levels as captured in Figure 3(c). Thus these samples do not exhibit the same wide programmability window seen in other devices. Nevertheless, they exhibit a steeper slope, just as TiO$_x$ devices do when formed in lower resistive states.
One notable observation that came as a result of changing the device area is that by electroforming, the devices do not lose their capacitive characteristics. Thus forming and subsequent switching in low resistive states does not lead to shorting the capacitor.

\begin{figure}[t]
    \centering
    \includegraphics[width = 9cm, height = 7cm]{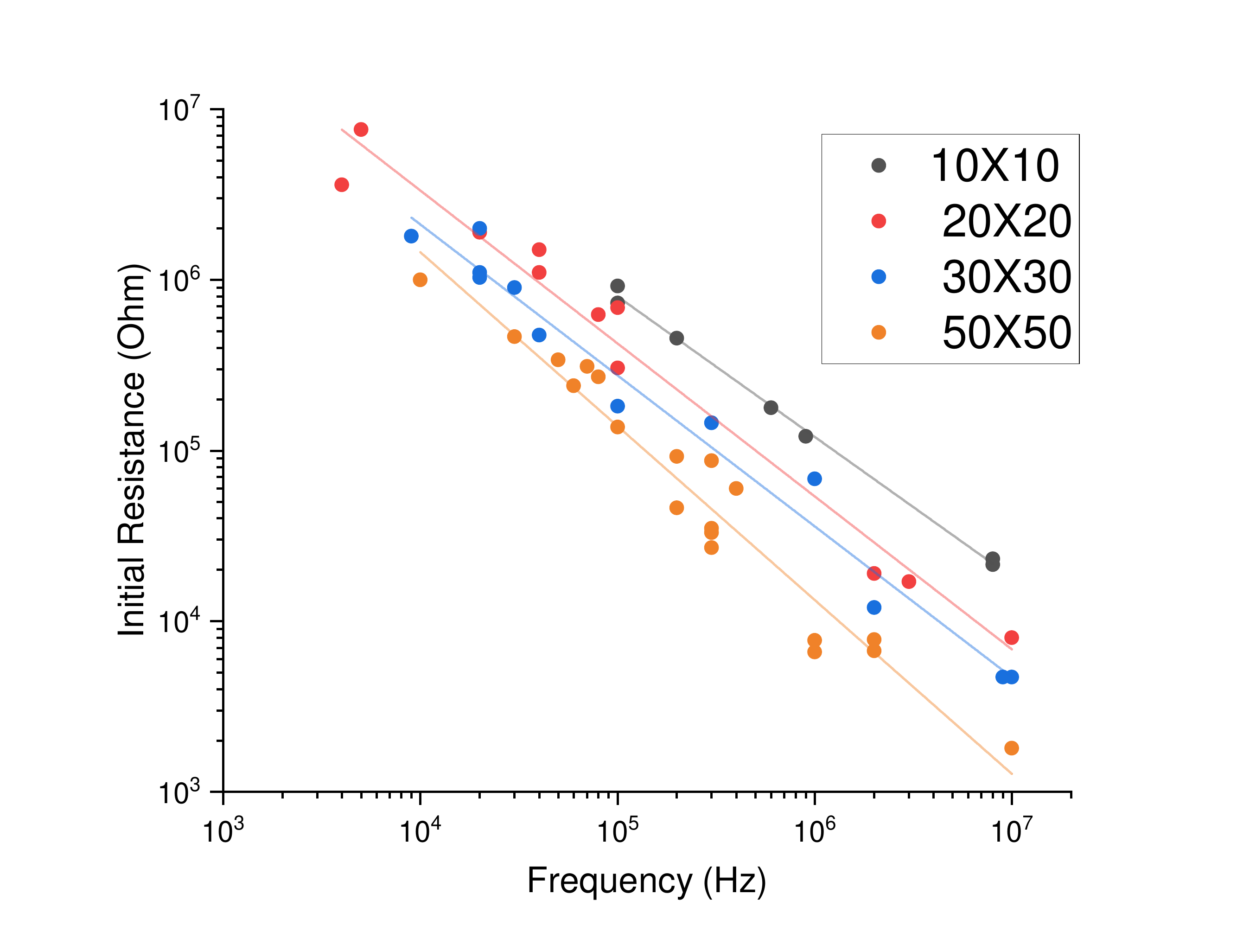}
    \caption{TiO$_x$ devices formed in low resistance skew the slope downwards indicating a possible dielectric change which could explain this change of behaviour}
    \label{fig 4}
\end{figure}

\subsection{Variation of dielectric}

Through fabricating devices with different dielectric layers the intent was to, firstly, confirm that the frequency response of such devices does not change and observe what overall effect, if any, that change in dielectric constant will have on the devices.

The switching behavior for devices with a single TiO$_x$ active core, followed the characteristics observed in TiO$_x$/Al$_2$O$_3$ devices for initial resistances of over approximately 20 kΩ. Nonetheless, when such memristors were formed around the 20 kΩ level, differences in behavior were observed. In this case, the cut-off slope becomes steeper, thus indicating that a significant change in the dielectric of the devices has transpired. Figure 4 illustrates the cut-off frequencies plotted against the initial resistance of TiO$_x$ devices. Here, we have also added devices that formed on lower resistances for comparison. When comparing Figure 4 to Figure 3(b) it is evident that the slope becomes steeper for lower resistances. This could be attributed to the potential appearance of a second , steeper, cut-off or a change in the dielectric. Similar observations for this specific device morphology were made by Michalas et al \cite{Michalas}, indicating that TiO$_x$ devices formed in specific resistive ranges exhibit a distinct behavior from other devices formed in higher resistance regimes. It is possible that irreversible alteration to the dielectric could lead to this behavior and this could mean a morphological change in the dielectric itself. For all other cases, where the dielectric does not change in a measurable and irreversible way, the behaviour is consistent.

Another observation is the deviation from a linear trend. Devices with a TiO$_x$/Al$_2$O$_3$ core stack, show a stricter adherence to a linear fit. TiO$_x$ devices on the other hand show a bigger margin of linearity deviation. This could be attributed to the forming process once again, as these deviations can be attributed to small deviations from expected capacitance for a given device area. It is possible that electroformed devices through the added Al$_2$O$_3$ layer may lead to a more streamlined and predictable device behaviour, with more gentle changes to the dielectric, in comparison with devices comprising a single-layer core. This is also supported by findings in earlier publications \cite{Stathopoulos2} \cite{Jeon}.

\begin{figure}
    \centering
    \includegraphics[width = 9cm, height = 7cm]{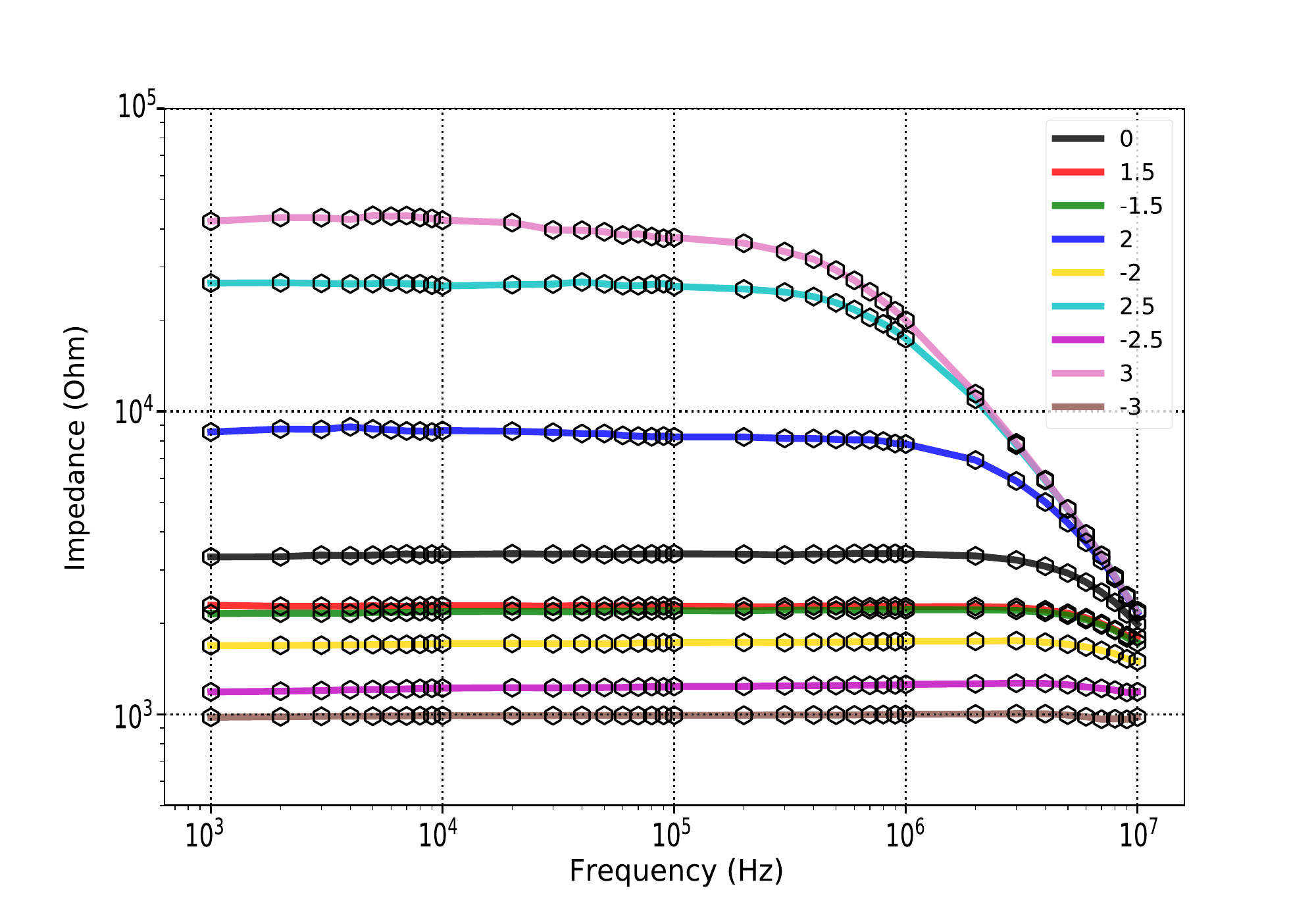}
    \caption{Devices with SnO$_x$ dielectric form in lower resistive ranges and exhibit repeatable switching. High resistive values lead to instability with applied frequency, which is only exacerbated in higher resistances. Legend shows the maximum voltage reached in each switching sweep.}
    \label{fig 5}
\end{figure}

\begin{figure}[h!]
    \centering
    \includegraphics[width = 9cm, height = 7cm]{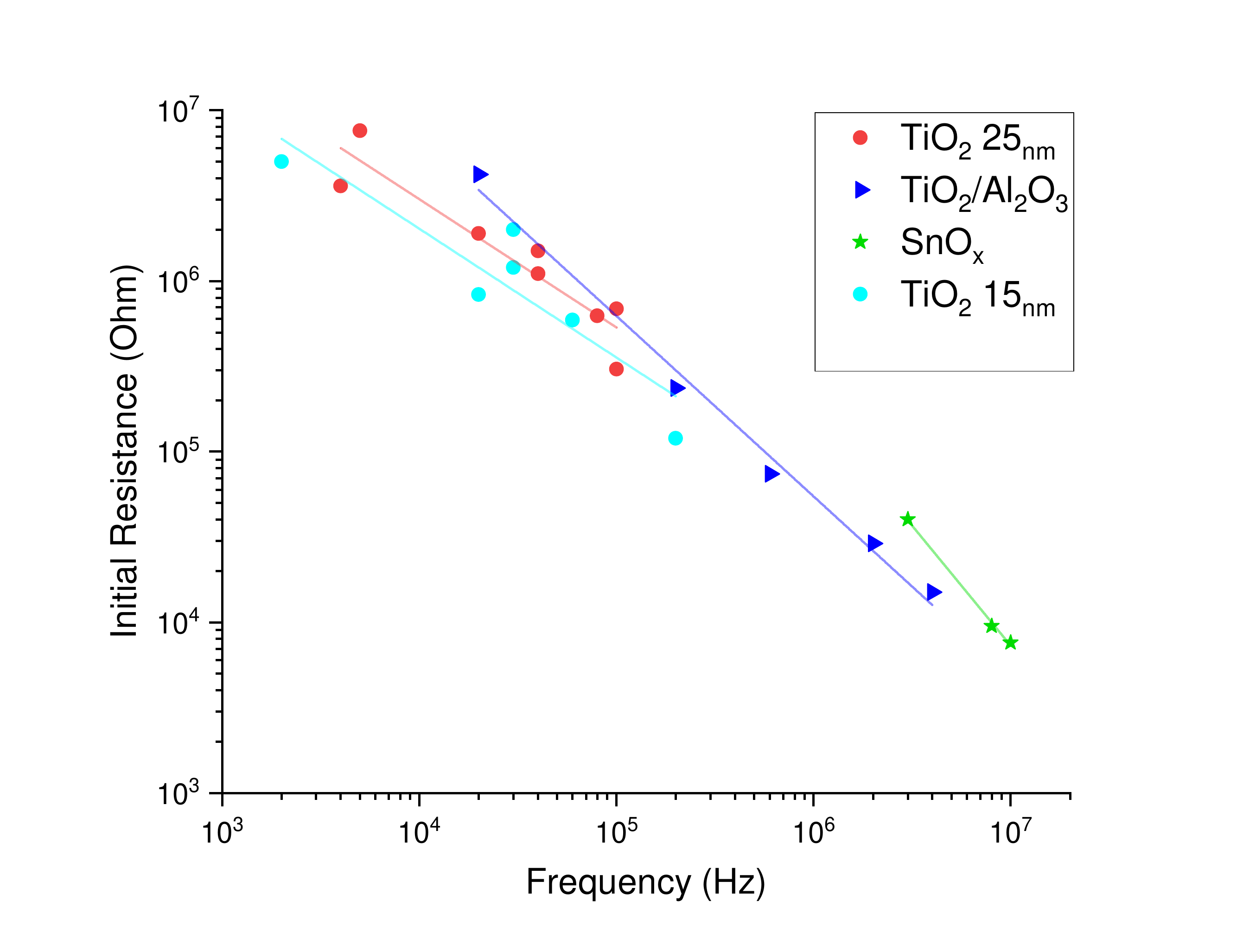}
    \caption{Cutoff slopes of devices with different dielectric. All devices had a 20$\times$20 um electrode area. The results depicted in Cyan are from devices with TiO$_x$ dielectric but a smaller dielectric thickness of 15nm}
    \label{fig 6}
\end{figure}

SnO$_x$ devices exhibit an overall steeper cut-off slope. Otherwise, their switching behaviour remains comparable to the other devices presented herein, meaning that semiconducting oxides as an active layer do not have an impact on it. One apparent change is that their impedance appears to be unstable in high resistive states, as observed in Figure 5; a characteristic also observed when running a retention test on devices at relatively high resistive states. 

Stability in the low resistive state follows the behaviour of TiO$_x$/Al$_2$O$_3$ devices. The studied behaviour of devices with regards to the stimuli frequency could potentially help when trying to link different conduction mechanisms with devices that have different characteristics, as already shown for TiO$_x$ devices. Along this line, Figure 6 displays a compilation of the cut-off slopes for devices with 20$\times$20 μm$^2$ area and different dielectric.

\begin{figure}[b]
    \centering
    \includegraphics[width = 9cm, height = 7cm]{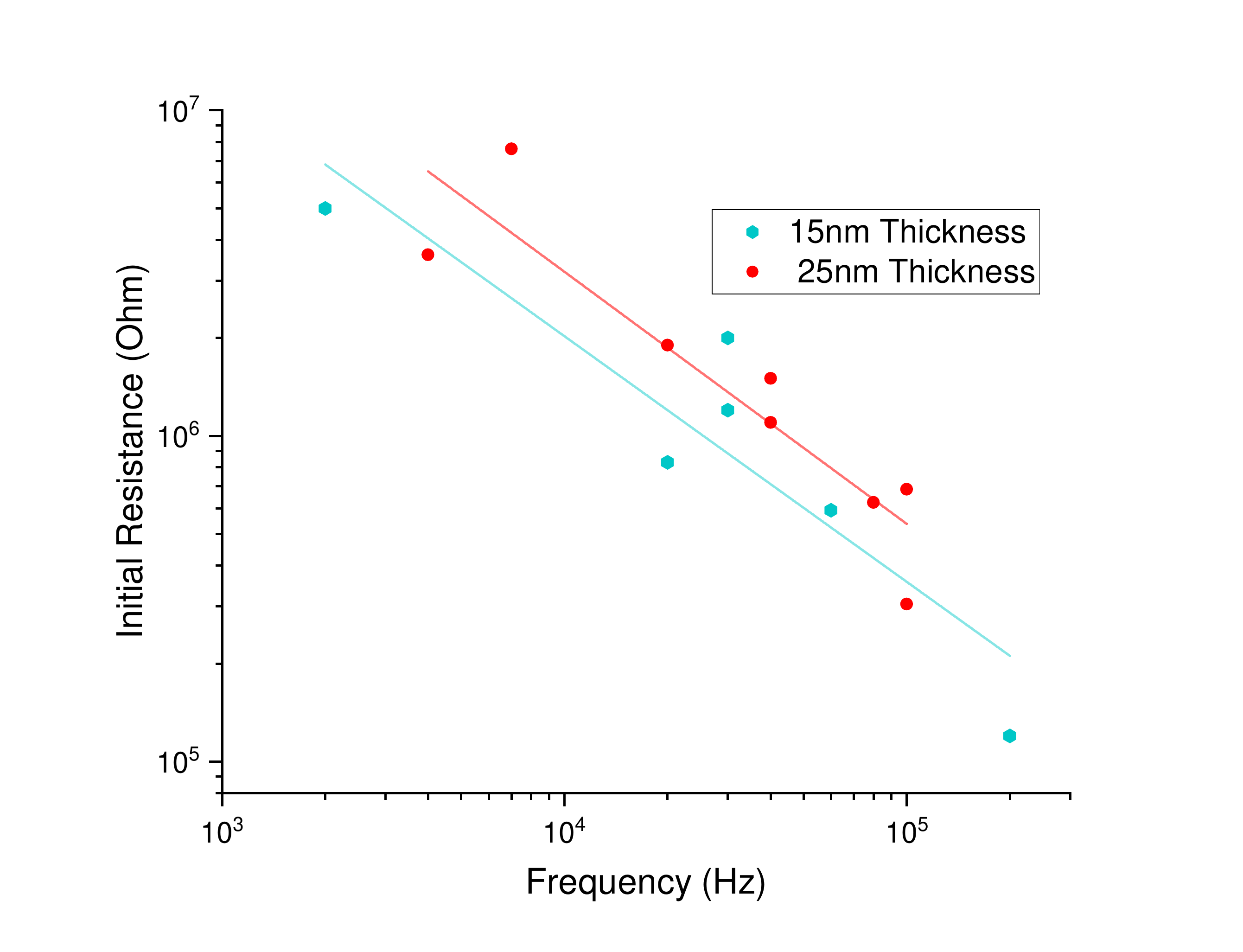}
    \caption{A change in dielectric thickness results in a higher capacitance, thus altering the programmability window of the device and offering another way to tailor device behavior to the needs of prospective applications}
    \label{fig 7}
\end{figure}

\subsection{Variation by changing dielectric thickness}

To investigate the effect of thickness in device behaviour, memristive cells with thinner dielectrics were also fabricated and tested. As expected, this increases the overall effective capacitance of the cells. Devices with 15 nm of TiO$_x$ dielectric were fabricated and their behaviour and cut-off was compared to the original TiO$_x$ devices, which had a 25 nm thick dielectric. Oxygen content and thickness of the dielectric are considered crucial parameters, which have an active role in device behavior \cite{Regoutz}. Generally, changes in these two parameters affect the ability to electroform the devices consistently. Nevertheless, after being electroformed, their frequency characteristics are consistent with what has already been discussed in previous sections. In Figure 7, a comparison between devices with 25 nm thick dielectric and devices with 15 nm thick dielectric is given. As expected in a parallel plate capacitor, a decrease in dielectric thickness leads to an increase in capacitance, thus making the cut-off slope move towards lower frequencies, in line with the simulated results in Figure 3(d). Thus, by changing dielectric thickness, the expected characteristics of devices can be fine tuned.

\section{Conclusion}

In this paper, we presented results from the behavior of Metal Oxide memristive devices in alternating current conditions, for a range of frequencies from $10^3$ up to $10^7$ Hz. The electrical programmability of resistive switching carries over into the frequency range tested here with the devices presenting an RC low-pass filter-like behavior, with the cut-off frequency of the device being tuned by its resistive state. 

This study expanded to account for most common device physical parameters, such as, area, dielectric material and dielectric thickness; a large range of samples were fabricated and characterized. It was found that changing parameters affecting the device's capacitance, such as device area and dielectric thickness, corresponds to a modulation of the frequency response via transferring the entire impedance plot parallel to the frequency axis.  This was consistent with an RC empirical model description, as shown through the excellent agreement between simulated and measured data for different capacitance values. This lends credence to the initial hypothesis that these parameters could be used to control the window of programmability by disproportionately altering capacitance, while leaving resistance of devices largely untouched. 

A change in dielectric leads to a distinct capacitance value, while also changing the slope of the impedance plot, thus the impact from different dielectrics is a key parameter on setting the frequency response of tunable cells, offering opportunities to tailor this for specific application. There are indications that other underpinning parameters which impact the DC characteristics of the devices, such as conduction mechanism, may have an effect in AC conditions, as for example shown for HRS states in SnO$_x$ devices and for TiO$_x$ devices formed in a lower resistive range. The selection of device area and dielectric thickness can both be used as a secondary parameter that allows tailoring the frequency response of said device in accordance to the applications' needs. While area can be changed with minimal impact on device behaviour, the same can not be said for dielectric thickness, where a change might make the electroforming process harder, thus the preferred parameter to change is electrode area.

\ifCLASSOPTIONcaptionsoff
  \newpage
\fi

\printbibliography

\end{document}